\documentclass[aps,prd,twocolumn,preprintnumbers,amsmath,amssymb,floatfix,nofootinbib]{revtex4-1}

\usepackage{aas_macros}
\usepackage{graphicx}

\usepackage{bm}
\usepackage[colorlinks=true]{hyperref}

\begin{document}


\title{Emergent gravity in galaxies and in the Solar System}

\author{Aur\'elien Hees}\email{ahees@astro.ucla.edu} \affiliation{Department of Physics and Astronomy, University of California, Los Angeles, CA 90095, USA}%

\author{Benoit Famaey} 
\affiliation{Universit\'e de Strasbourg, CNRS, Observatoire astronomique de Strasbourg,
UMR 7550, F-67000 Strasbourg, France}%

\author{Gianfranco Bertone}
\affiliation{GRAPPA, University of Amsterdam, Science Park 904, 1098 XH Amsterdam, Netherlands}%

\date{\today}

\begin{abstract}
It was recently proposed that the effects usually attributed to particle dark matter on galaxy scales are due to the displacement of dark energy by baryonic matter, a paradigm known as {\it emergent gravity}. This formalism leads to predictions similar to Modified Newtonian Dynamics (MOND) in spherical symmetry, but not quite identical. In particular, it leads to a well defined transition between the Newtonian and the modified gravitational regimes, a transition depending on both the Newtonian acceleration and its first derivative with respect to radius. Under the hypothesis of the applicability of this transition to aspherical systems, we investigate whether it can reproduce observed galaxy rotation curves. We conclude that the formula leads to marginally acceptable fits with strikingly low best-fit distances, low stellar mass-to-light ratios, and a low Hubble constant. In particular, some unobserved wiggles are produced in rotation curves because of the dependence of the transition on the derivative of the Newtonian acceleration, leading, even in the most favorable case, to systematically less good fits than MOND. Then, applying the predicted transition from emergent gravity in a regime where it should a priori be applicable, i.e. in  spherical symmetry and outside of the bulk of matter, we show that the predictions for the secular advances of Solar System planets' perihelia are discrepant with the data by seven orders of magnitude, ruling out the present emergent gravity weak-field formula with high confidence.
\end{abstract}

\maketitle

\section{Introduction} 

There is overwhelming evidence that astrophysical and cosmological observations on scales of dwarf galaxies and above cannot be explained, in the framework of general relativity, in terms of known elementary particles \cite{Bertone:2010zza,Sanders:2010cle,Bertone:2016nfn}.
In the standard cosmological model, the problem is solved by postulating the existence of a new matter component, dubbed {\it dark matter}, which is made of new, yet undiscovered particles, possibly connected with extensions of the Standard Model of particle physics proposed for completely independent reasons, such as WIMPs (weakly interacting massive particles), axions, and sterile neutrinos \cite{Jungman:1995df,Bergstrom:2000pn,Bertone:2004pz,Feng:2010gw,Klasen:2015uma}. 

Despite a large theoretical and experimental effort, however, there is no conclusive evidence for the existence of any of those dark matter candidates. Furthermore, our current models of galaxy formation in the standard cosmological model appear to struggle to reproduce observations \citep[e.g.][]{pawlowski:2015aa}. This might be due to hard-to-model feedback processes linked to the physics of baryons in galaxies \citep[e.g.][]{Frenk:2012ph, Read:2016aa}, or to the fact that the properties of dark matter particles, and especially their interactions with themselves \cite{Spergel:1999mh} (see also \cite{Kamada:2016euw} and references therein) or with baryons \citep[e.g.][]{Berezhiani:2015aa}, might be different from those currently assumed in standard cosmology.

As long appreciated \cite{1963MNRAS.127...21F}, an alternative solution to explain cosmological observations is to introduce a suitable modification of  the laws of gravity \citep[e.g.][]{milgrom:1983fk,*milgrom:1983kx,*milgrom:1983uq,Bekenstein:2010pt}. The case for such an effective modification of gravity is best summarized by the fact that observed galaxy rotation curve shapes are diverse at a given maximum velocity scale \citep{oman:2015aa}, where they are expected to be uniform, and uniform at a given baryonic surface density scale \citep{famaey:2012fk,lelli:2013aa}, where they are expected to be diverse. The latter means that exponential disks of the same baryonic mass have similar rotation curves only if the radius is renormalized by the disk scale-length \citep{mcgaugh:2014aa}. Such a behaviour, which appears {\it a priori} unnatural in the standard context, has long been predicted by Modified Newtonian Dynamics \citep[MOND,][]{milgrom:1983fk,*milgrom:1983kx,*milgrom:1983uq}, which posits that  
\begin{equation}\label{eq:mond}
\bm g = \nu\left(\frac{g_\mathrm{N}}{a_M}\right) \bm g_\mathrm{N},
\end{equation}
where $\bm g$ and $\bm {g}_\mathrm{N}$ are the effective and Newtonian gravitational accelerations respectively, $g_N=\left|\bm g_N\right|$ and $\nu$ is a transition function verifying
\begin{equation}
\left\{\begin{array}{ll}
\nu (x) \rightarrow 1 & \textrm{for}\quad x \gg 1 \\
\nu (x) \rightarrow x^{-1/2} &\textrm{for}\quad x \ll 1 \\
\end{array}
\right..
\end{equation}
This relation involves an acceleration scale $a_M \sim c^2/L \sim 10^{-10} {\rm m} \, {\rm s}^{-2}$ where $L$ is the Hubble scale\footnote{We adopt here notations consistent with those used in the emergent gravity framework. In the MOND context, the acceleration constant is rather usually denoted as $a_0$.}, also roughly corresponding to the curvature radius corresponding to the $z=0$ {\it dark energy} content of the Universe. Note that, outside of spherical symmetry, this formula cannot be correct and classical MOND theories rather modify the Poisson equation, following the stationary action principle for a modified Lagrangian of gravitation involving this acceleration constant \citep{famaey:2012fk}, which allows us to make detailed simulations outside of spherical symmetry \citep[e.g.,][]{lughausen:2015jt, renaud:2016aa}. Nevertheless, for galaxy rotation curves fits, the difference between the pristine formula from Eq.~(\ref{eq:mond}) and the actual modification of gravity prediction is small enough \citep{angus:2012cs}, so that Eq.~(\ref{eq:mond}) is commonly used to predict galaxy rotation curve shapes.

Inspired by the success of the MOND formalism on galaxy scales, a framework in which a similar relation naturally emerges was recently proposed \citep{verlinde:2016qf}. The core idea dates back to the realization by Milgrom \citep{milgrom:1999aa} that a redefinition of inertia as being proportional to the vacuum temperature seen by an accelerated observer could naturally lead to the MOND relation, since such an observer in a de Sitter universe sees a non-linear combination of the Unruh vacuum radiation and of the  Gibbons-Hawking radiation due to the  cosmological horizon. A slightly more fundamental (albeit currently not really full-fledged, see \cite{milgrom:2016aa}) approach was then proposed by Verlinde \citep{verlinde:2016qf}, in which gravity emerges from the entanglement entropy of the vacuum. In this picture, de Sitter space corresponds to a set of metastable quantum states carrying the entropy associated with the cosmological horizon, and the positive dark energy is caused by the slow thermalization of the emergent spacetime. In the presence of baryonic matter, this dark energy is slightly ``displaced" which leads to an effective modification of gravity in the low acceleration regime, below $\sim c^2/L$. In this regime, the apparent dark matter mass $M_D$ appears as directly related to the baryonic one. In spherical symmetry, this formalism leads to predictions that are essentially similar to those of MOND, but not quite identical. In particular, it predicts a well-defined transition between the classical and the low-acceleration modified gravitational regimes, a transition depending on both the Newtonian acceleration and its first derivative with respect to radius. This formalism has already been successfully confronted with galaxy-galaxy lensing data \citep{brouwer:2017aa}, distance luminosity and baryon acoustic oscillations \citep{zeng:2017aa} and dwarf spheroidal galaxies \citep{diez-tejedor:2016aa}, and has encountered some problems in galaxy clusters \citep{ettori:2016aa}, akin to those of MOND at these scales \citep{sanders:1999aa, angus:2008aa}.

Here, we set out to estimate the generic consequences of emergent gravity for the predictions of galaxy rotation curve shapes, with the caveat that the formalism might not be fully applicable {\it per se} outside of spherical symmetry, and we also quantify criticisms already made at the qualitative level \citep{milgrom:2016aa} regarding its predictions in the Solar System, where the predictive formula from emergent gravity should be fully applicable.

\section{The MOND and emergent gravity transitions}

In MOND, neither the exact value of the acceleration constant nor the shape of the transition function are hardwired into the paradigm. Hence there is a little bit of freedom in choosing those to best reproduce galaxy rotation curves. In \citet{hees:2016mi}, we combined galaxy rotation curves and Solar System data to put constraints on the transition function $\nu$, and showed that one class of functions allowed us to fit rotation curves while escaping Solar System constraints\footnote{Note that the empirically successful transition in Equation~(4) of \cite{mcgaugh:2016aa} is excluded by Solar System constraints in modified gravity versions of MOND, as it corresponds to $\hat \nu_1$ in Tab.~2 of \cite{hees:2016mi}.}. One function which is representative of this family is
\begin{equation}\label{eq:nuMOND}
\nu_{\rm MOND}(x) =  \left(1-e^{-x^2}\right)^{-1/4}+\frac{3}{4}e^{-x^2},
\end{equation}
and has been shown to reproduce galaxy rotation curves well, especially when including the contribution of the {\it external field effect} \citep{hees:2016mi}.

In emergent gravity, the apparent dark matter mass $M_D$ is directly related to the baryonic mass $M_B$ (see Eq.~(7.40) from \citep{verlinde:2016qf}) through
\begin{equation}
	\int_0^r \frac{GM^2_D(\tilde r)}{\tilde r^2}d\tilde r=M_B(r) a_M r \, ,
\end{equation}
with $a_M$ a constant acceleration scale in principle  accurately defined as $a_M = c^2/6L = cH_0/6$. Within this theory, the total gravitational acceleration can be written in a MOND-like form using the following transition function 
\begin{equation}\label{eq:nuEG}
\nu_{\rm EG}(x, x', r) =  1 + \sqrt{\frac{3}{x} + \frac{x' r}{x^2}} \, ,
\end{equation}
which depends explicitly on the radius $r$ and on the radial derivative $x'$ of the Newtonian gravitational acceleration.

In the following sections, we will quantify the observational consequences of this transition function in galaxies and the Solar System. One aspect of emergent gravity which is not completely clear is whether it would lead to an external field effect similar to that of MOND. In the next section, we will show the predictions for galaxy rotation curves with and without such an effect.

\section{Rotation curve fits}
In this section, we produce traditional fits to rotation curves~\citep{begeman:1991fk,sanders:1998uq,de-blok:1998zr,sanders:2007ly,gentile:2011uq,hees:2016mi} using the emergent gravity paradigm described by the transition function from Eq.~(\ref{eq:nuEG}). We have to assume here that, as is the case in MOND, the predictions outside of spherical symmetry do not deviate too much from the spherically symmetric solution. Three different scenarios are then considered: (i) a scenario where the value of the  acceleration scale $a_M=cH_0/6$ is fixed by using existing local measurements of the Hubble constant~\citep{riess:2016nr} ; (ii) a scenario where the acceleration scale $a_M$ is left free ;  and (iii) a scenario where the external field effect is also considered. In order to compare our results with results obtained with the MOND phenomenology, we also produce a fit using the MOND transition function from Eq.~(\ref{eq:nuMOND}).

We use the same rotation curve data as in \cite{hees:2016mi}, for 27 dwarf and low surface brightness galaxies thoroughly described in \citet{swaters:2010eu}. Our fit includes one global parameter, the acceleration scale $a_M$, and two local galactic parameters: the individual $R$-band stellar mass-to-light (M/L) ratio $\Upsilon_g$ and a relative rescaling of the distance to the different galaxies $d_g$ (the indices $g$ refer to a particular galaxy and indicate that the parameters are local parameters) within the errors of the estimated distance. In addition, the third scenario includes an additional external Newtonian field $g_{Neg}$ for each galaxy.

The predicted rotation velocity $V$ at radius $R_i$ is given by
\begin{equation}
	V(R_id_g;a_M,\Upsilon_g,d_g)=\sqrt{R_i d_g g(R_i;a_M,\Upsilon_g)} \, ,
\end{equation}
where $\Upsilon_g$ is the stellar M/L ratio, $d_g=D_{g}/D_{g,0}$ where $D_{g}$ is the distance used here and $D_{g,0}$ the distance given in Tab.~1 of \citep{swaters:2010eu}. The norm of the gravitational field $g$ is determined by Eq.~(\ref{eq:mond}) with the transition function from Eq.~(\ref{eq:nuEG}), and the Newtonian gravitational field can be expressed in terms of the Newtonian velocities as
\begin{equation}
	g_N(R_i,\Upsilon_g)=\frac{V^2_{{\rm gas}i}}{R_i}+\Upsilon_g\frac{V_{\star i}^2}{R_i} \, ,
\end{equation}
where $V_{{\rm gas}i}$ and $V_{\star i}$ are the contribution of the gas and of the stellar disk (at radius $R_i$) to the rotation curves calculated in the Newtonian framework without dark matter. In what precedes, we have used the fact that the Newtonian observed velocities due to the gas and to the stellar disk are rescaled as $\propto \sqrt{d}$ with a distance rescaling. Similarly, the measured radial distances $R_i$ are rescaled proportionally to $d$. The transition function from Eq.~(\ref{eq:nuEG}) requires an estimation of the derivative of $g_N$ trivially given by
\begin{equation}
	g'_N(R_i,\Upsilon_g)=-\frac{g_N(R_i,\Upsilon_g)}{R_i} +2\frac{V_{{\rm gas}i}V'_{{\rm gas}i}}{R_i} + 2\Upsilon_g \frac{V_{\star i}V'_{\star i}}{R_i}\, .
\end{equation}
Note that the quantity $R_i g'_N$ appearing in the expression of the transition function from Eq.~(\ref{eq:nuEG}) is independent of the rescaling factor $d$. 

The procedure to analyze the data is identical to the ones presented in details in \citep{hees:2016mi}. In a first step, a least-square fit of the global acceleration scale $a_M$ and of the local $\Upsilon_g$ and $d_g$ parameters is performed using a subset of 19 galaxies. The galaxies not considered in this step have been identified to experience an hypothetical external field effect that may bias the estimation of $a_M$ (see the discussion in \citep{hees:2016mi}). In a second step, using the optimal value of $a_M$ previously obtained, we perform a Bayesian inference for the parameters $\Upsilon_g$ and $d_g$ (the parameter $g_{Neg}$ is included in the third scenario). During the analysis, we always impose the scaling of the distance to be between 0.7 and 1.3 and the stellar M/L ratios to have values included between 0.3 and 5 (in units of $(M/L)_{\odot}$). Furthermore, we also use a Gaussian prior (characterized by a mean of 1 and a standard deviation of 0.1) on the parameters $d_g$.

\subsection{Using a fixed value of the Hubble constant}\label{sec:prior}
\begin{figure*}
\includegraphics[width=1.7\columnwidth]{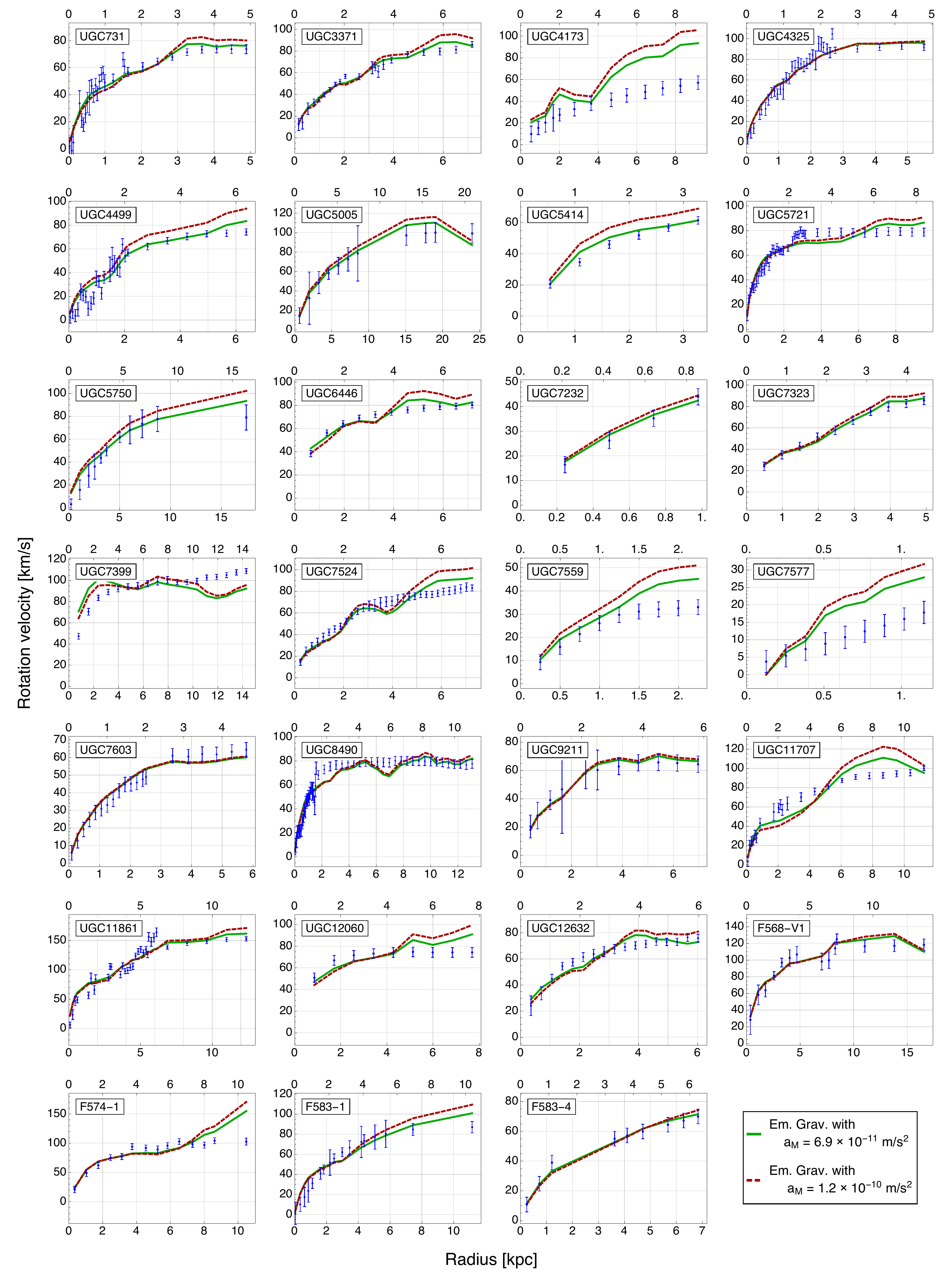}
\caption{Results of the fit using the  emergent gravity ($\nu_\textrm{EG}$) without any external field effect. The red, dashed curves correspond to a fit using a fixed value for $a_M$ motivated from local measurements of $H_0$ \citep{riess:2016nr} while the continuous green curves are related to a fit without any prior on the $a_M$ parameter. Since the optimal fits do not necessarily produce the same distance scale factor, the radial scales may not be the same. On the top of the plots we mention the radial scale obtained with the prior (corresponding to the dashed red  lines), at the bottom of the plots we mention the radial scale obtained without any prior (corresponding to the  green solid lines).}
\label{fig:resnoEFE}
\end{figure*}

In emergent gravity, the acceleration scale $a_M$ is directly related to the Hubble constant through $a_M=cH_0/6$. In this section, we fix the value of the acceleration scale by using the estimation $H_0=73.24$ km/s/Mpc obtained from local galaxy distance scales in \citep{riess:2016nr}, which corresponds to  $a_M=1.2\times 10^{-10}$~m/s$^2$.  The optimal values and the 68\% confidence intervals for the local parameters $\Upsilon_g$ and $d_g$ are presented in Tab.~\ref{tab:res}. For the majority of the galaxies, the distance scale factor is lower than unity, indicating that within the emergent gravity paradigm, all galaxies should be much closer than their current observational estimate. The reduced chi-square for the fit including the 27 galaxies is equal to 7.1, i.e. unacceptably high. The red, dashed curves in Fig.~\ref{fig:resnoEFE} show the optimal rotation curves obtained for each galaxy. It is interesting to note that emergent gravity produces too large wiggles in some galaxy rotation curves: UGC7524, UGC8490, UGC 11707 are the most striking cases. This feature is due to the contribution arising from the derivative of the Newtonian gravitational acceleration in the transition function from Eq.~(\ref{eq:nuEG}). When wiggles exist in the Newtonian rotation curves, the derivative term tends to artificially amplify those in the predicted emergent gravity rotation curves. The conclusion from this fit is that using a value of $a_M$ coming from local measurements of the Hubble constant is incompatible with observations of galactic rotation curves. Therefore, in the following section, we will relax this assumption.

\subsection{No prior on the acceleration scale}\label{sec:noprior}
One can argue that the value of the Hubble constant inferred from different data in the context of emergent gravity may  differ from its estimation assuming General Relativity. Alternatively, some small tweaks in the emergent gravity formalism could perhaps change the $1/6$ factor in some smaller factor, while not changing the general formula. This could for instance be a conceivable consequence of moving from spherical symmetry to axisymmetry. If this would be the case, the value of $a_M$ used in the previous section may bias the result. Therefore, we produce a global fit more favorable to emergent gravity, where the acceleration scale is left as a free parameter. First, we use the local estimate of the Hubble constant and its uncertainties as a prior, and get a best-fit value $a_M=8.6 \times 10^{-11}$~m/s$^2$ corresponding to $H_0=6 a_M/c =$52.9~km/s/Mpc and a reduced chi-square of 5.1. Then, we let the parameter $a_M$ as completely free (no prior). The optimal acceleration scale is then given by $6.9 \times 10^{-11}$~m/s$^2$, which corresponds to an Hubble constant of $H_0=42.6$ km/s/Mpc. This value is significantly lower than the ones obtained from local measurements \citep{riess:2016nr} or from Planck observations \citep{planck-collaboration:2016kx}. 

The reduced chi-square for this fit with no prior on $a_M$ is 4.4, showing an improvement with respect to the fit presented in the previous section. The optimal values and the 68\% confidence intervals for the local parameters $\Upsilon_g$ and $d_g$ are presented in Tab.~\ref{tab:res}. Globally, these values are slightly higher than the corresponding ones obtained in the previous section, which is coming from a correlation with $a_M$, but they are still much lower than expected on average. The optimal rotation curves are the green, solid curves displayed on Fig.~\ref{fig:resnoEFE}. Several curves are marginally improved with respect to the fit produced using a prior on $a_M$, like for instance UGC4173, UGC4499, UGC5414, UGC5721, UGC7559, or UGC7577. The inconvenient wiggles are still present (UGC7524, UGC8490, UGC 11707).

\subsection{Including an external field effect}\label{sec:EFE}
Within the MOND paradigm, it is known that the external field in which the system is embedded impacts the local gravitational dynamics  \citep{milgrom:1983fk}. This external field effect which appears even for a constant external field is due to the non-linearity of the MOND theory and is a consequence of a violation of the strong equivalence principle. While not mentioned in \cite{verlinde:2016qf}, and not taken into account in, e.g., \cite{brouwer:2017aa}, a similar effect may in principle arise in emergent gravity. This would be the most favorable situation possible, as it should in principle allow to significantly improve the fits, if combined with a low value of $a_M$.

Decomposing the total gravitational field into an internal part $\bm g$ and an external field $\bm g_e$ and using a similar decomposition for the Newtonian gravitational acceleration ($\bm g_N+\bm g_{Ne}$) allows us to generalize Eq.~(\ref{eq:mond}) taking into account the external field contribution. As in \cite{hees:2016mi}, we can use as first approximation the one-dimensional version of Eq.~(\ref{eq:mond}), which becomes
\begin{align}
	g&=\nu_\textrm{EG}\left( \frac{g_N+g_{Ne}}{a_M}, \frac{g_N'}{a_M},R\right)\left( g_N +  g_{Ne}\right)\\
	  &\qquad\qquad-\nu_\textrm{EG}\left(\frac{g_{Ne}}{a_M},0,R\right) g_{Ne}\, , \nonumber
\end{align}
where we assume the external gravitational field to be constant over the system.

\begin{figure*}
\includegraphics[width=1.7\columnwidth]{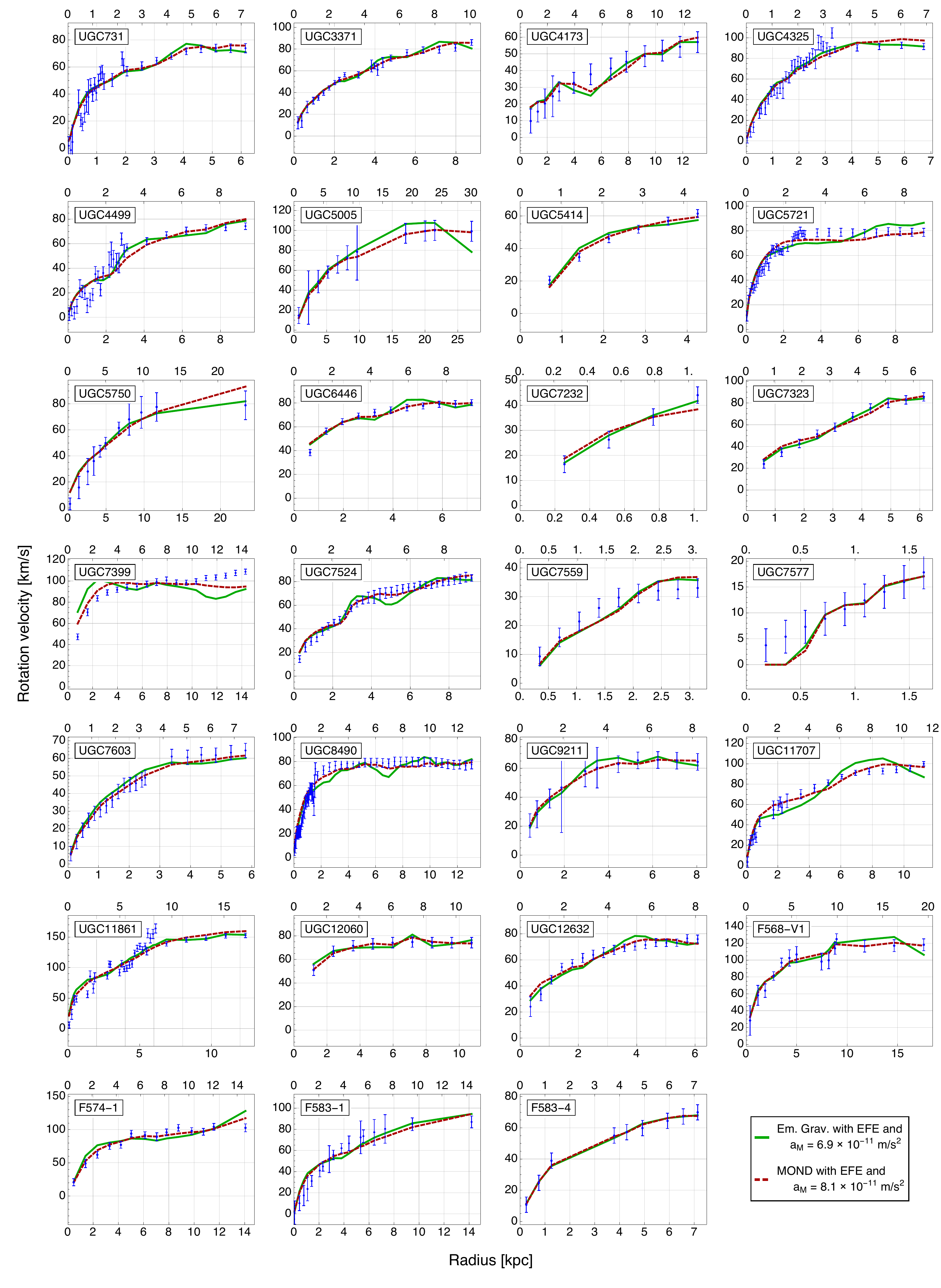}
\caption{Results of the fit using the  emergent gravity ($\nu_\textrm{EG}$) and the MOND transition function $\nu_\textrm{MOND}$ including the external field effect. Since the optimal fits do not necessarily produce the same distance scale factor, the radial scales may not be the same. On the top of the plots we mention the radial scale obtained with MOND (corresponding to the dashed red thick lines), at the bottom of the plots we mention the radial scale obtained with the emergent gravity (corresponding to the thick green solid lines).}
\label{fig:rotEFE}
\end{figure*}

We produce a fit including a third local parameter for each galaxy: the value of the external gravitational field $g_{Ne}$. The reduced chi-square for the fit is now 3.5. Including the external field effect thus improves significantly the quality of the fit. The optimal rotation curves obtained for this fit are presented as the green solid line in Fig.~\ref{fig:rotEFE}. As thoroughly discussed in \citep{hees:2016mi}, the external field effect improves significantly the outer part of the fitted rotation curves for several galaxies like e.g. UGC4173, UGC499, UGC7524, UGC7559, UGC7577,  UGC12060 and F574-1. 

Nevertheless, for the sake of comparison, we also produce (red dashed line on Fig.~\ref{fig:rotEFE}) a similar fit with the MOND transition function from Eq.~(\ref{eq:nuMOND}).  The reduced chi-square for this MOND fit is 2.1, a traditionally acceptable value for rotation curves, due to somewhat underestimated systematics in error bars. Fig.~\ref{fig:rotEFE} clearly shows that several fits with emergent gravity are significantly worse than using the MOND transition function. The main reason is the presence of wiggles in some fits (e.g.  UGC7524, UGC8490, UGC 11707) which deter significantly the quality of the fit. These wiggles are due to the presence of the derivative of the Newtonian gravitational acceleration in the emergent gravity transition function from Eq.~(\ref{eq:nuEG}). The local optimal parameters ($\Upsilon_g$, $d_g$ and the external gravitational field $g_{Neg}$) and their corresponding 68\% confidence interval are given in Tab.~\ref{tab:res}. The emergent gravity produces fits with smaller stellar mass-to-light ratios and systematically low distance scale factors.

\begin{table*}[htb]
	\caption{Best-fit local parameters obtained for the different scenarios considered in this analysis: emergent gravity using a fixed value for $a_M$ (see Sec.~\ref{sec:prior}), emergent gravity with $a_M$ as a free parameter (see Sec.~\ref{sec:noprior}), emergent gravity with an external field effect (see Sec.~\ref{sec:EFE}) and the MOND transition function from Eq.~(\ref{eq:nuMOND}) and the external field effect. The values reported are optimal values and 68 \% Bayesian confidence intervals for the parameters. For $g_{Neg}$, only the values significantly different from 0 are reported.}
	\label{tab:res} 
	\centering
	\begin{tabular}{l|rr|rr|rrr|rrr}
	 & \multicolumn{2}{|c}{$\nu_\textrm{EG}$} & \multicolumn{2}{|c}{$\nu_\textrm{EG}$} &  \multicolumn{3}{|c}{$\nu_\textrm{EG}$} & \multicolumn{3}{|c}{$\nu_\textrm{MOND}$}\\
	& \multicolumn{2}{|c}{With fixed $a_M$} & \multicolumn{2}{|c}{No prior on $a_M$} &  \multicolumn{3}{|c}{With ext. field effect} & \multicolumn{3}{|c}{With ext. field effect} \\\hline 
	Opt. $a_M$ [m/s$^2$] & \multicolumn{2}{|c}{1.2 $\times  10^{-10}$ } & \multicolumn{2}{|c}{6.9 $\times  10^{-11}$ } & \multicolumn{3}{|c}{6.9 $\times  10^{-11}$ } & \multicolumn{3}{|c}{8.1 $\times  10^{-11}$}\\
	$\chi^2_\textrm{red}$ & \multicolumn{2}{|c}{7.1 } & \multicolumn{2}{|c}{4.4 } & \multicolumn{3}{|c}{3.5 }& \multicolumn{3}{|c}{2.1 } \\ \hline
  & $\Upsilon_g$ & $d_g$& $\Upsilon_g$ & $d_g$	& $\Upsilon_g$ & $d_g$ & $\log g_{Neg}$ & $\Upsilon_g$ & $d_g$ & $\log g_{Neg}$   \\
&$(M/L)_{\odot}$&&$(M/L)_{\odot}$&	&$(M/L)_{\odot}$&& (m/s$^2$) &$(M/L)_{\odot}$& &(m/s$^2$)\\\hline

	UGC731  &  $1.74_{-0.21}^{+0.11}$  &  $0.70_{-0.00}^{+0.01}$  &   $3.21_{-0.44}^{+0.11}$  &  $0.70_{-0.00}^{+0.02}$  &  $3.71_{-0.51}^{+0.54}$  &  $0.88_{-0.14}^{+0.06}$  &  $-11.80_{-0.42}^{+0.34}$  &  $5.0_{-0.49}^{+0.00}$  &  $1.03_{-0.19}^{+0.04}$  &  $-11.80_{-1.12}^{+0.44}$  \\ 
	UGC3371 &  $1.24_{-0.09}^{+0.06}$  &  $0.70_{-0.00}^{+0.01}$  &   $2.10_{-0.24}^{+0.03}$  &  $0.70_{-0.00}^{+0.02}$  &  $2.32_{-0.27}^{+0.24}$  &  $0.86_{-0.16}^{+0.00}$  &  $-12.20_{-0.78}^{+0.49}$  &  $3.3_{-0.56}^{+0.35}$  &  $0.99_{-0.19}^{+0.00}$  &  $-12.50_{-5.46}^{+0.70}$  \\ 
	UGC4173 &  $0.30_{-0.00}^{+0.02}$  &  $0.70_{-0.00}^{+0.01}$  &   $0.30_{-0.00}^{+0.04}$  &  $0.70_{-0.00}^{+0.01}$  &  $1.05_{-0.53}^{+0.22}$  &  $1.00_{-0.14}^{+0.05}$  &  $-10.40_{-0.21}^{+0.06}$  &  $0.8_{-0.38}^{+0.53}$  &  $0.99_{-0.12}^{+0.10}$  &  $-11.20_{-0.22}^{+0.37}$  \\ 
	UGC4325 &  $1.46_{-0.22}^{+0.21}$  &  $0.76_{-0.05}^{+0.04}$  &   $1.70_{-0.26}^{+0.33}$  &  $0.88_{-0.08}^{+0.06}$  &  $1.70_{-0.26}^{+0.28}$  &  $1.07_{-0.16}^{+0.07}$  &  $-11.80_{-0.62}^{+0.42}$  &  $3.7_{-0.60}^{+0.48}$  &  $1.10_{-0.10}^{+0.07}$  &  $-11.30_{-0.39}^{+0.26}$  \\ 
	UGC4499 &  $0.30_{-0.00}^{+0.01}$  &  $0.70_{-0.00}^{+0.01}$  &   $0.30_{-0.00}^{+0.02}$  &  $0.70_{-0.00}^{+0.01}$  &  $0.30_{-0.00}^{+0.02}$  &  $1.02_{-0.12}^{+0.08}$  &  $-11.70_{-0.24}^{+0.22}$  &  $0.3_{-0.00}^{+0.05}$  &  $1.03_{-0.11}^{+0.00}$  &  -                         \\ 
	UGC5005 &  $0.30_{-0.00}^{+0.09}$  &  $0.70_{-0.00}^{+0.03}$  &   $0.30_{-0.00}^{+0.19}$  &  $0.81_{-0.08}^{+0.03}$  &  $0.30_{-0.00}^{+0.20}$  &  $0.91_{-0.18}^{+0.00}$  &  $-13.20_{-4.93}^{+1.01}$  &  $0.9_{-0.56}^{+0.14}$  &  $1.01_{-0.16}^{+0.02}$  &  $-12.70_{-5.37}^{+0.76}$  \\ 
	UGC5414 &  $0.30_{-0.00}^{+0.01}$  &  $0.70_{-0.00}^{+0.01}$  &   $0.30_{-0.00}^{+0.03}$  &  $0.70_{-0.00}^{+0.01}$  &  $0.30_{-0.00}^{+0.24}$  &  $0.91_{-0.13}^{+0.11}$  &  $-11.90_{-0.28}^{+0.72}$  &  $1.0_{-0.55}^{+0.08}$  &  $0.93_{-0.18}^{+0.01}$  &  $-11.90_{-4.92}^{+1.60}$  \\ 
	UGC5721 &  $0.66_{-0.06}^{+0.06}$  &  $1.14_{-0.03}^{+0.03}$  &   $0.82_{-0.02}^{+0.05}$  &  $1.30_{-0.01}^{+0.00}$  &  $0.82_{-0.02}^{+0.06}$  &  $1.30_{-0.01}^{+0.00}$  &  -                         &  $2.4_{-0.25}^{+0.29}$  &  $1.23_{-0.04}^{+0.07}$  &  -                         \\ 
	UGC5750 &  $0.30_{-0.00}^{+0.03}$  &  $0.70_{-0.00}^{+0.02}$  &   $0.30_{-0.00}^{+0.07}$  &  $0.75_{-0.05}^{+0.02}$  &  $0.30_{-0.00}^{+0.14}$  &  $1.00_{-0.15}^{+0.07}$  &  $-12.30_{-0.32}^{+0.47}$  &  $0.3_{-0.00}^{+0.22}$  &  $1.02_{-0.13}^{+0.04}$  &  -                         \\ 
	UGC6446 &  $0.50_{-0.05}^{+0.04}$  &  $0.70_{-0.00}^{+0.01}$  &   $0.96_{-0.10}^{+0.04}$  &  $0.70_{-0.00}^{+0.01}$  &  $1.19_{-0.22}^{+0.03}$  &  $0.70_{-0.00}^{+0.07}$  &  $-13.00_{-0.06}^{+0.75}$  &  $1.6_{-0.14}^{+0.37}$  &  $0.91_{-0.21}^{+0.00}$  &  $-12.20_{-1.53}^{+0.62}$  \\ 
	UGC7232 &  $0.30_{-0.00}^{+0.09}$  &  $0.84_{-0.09}^{+0.04}$  &   $0.30_{-0.00}^{+0.17}$  &  $0.96_{-0.12}^{+0.04}$  &  $0.30_{-0.00}^{+0.25}$  &  $1.00_{-0.15}^{+0.03}$  &  $-13.00_{-5.00}^{+0.00}$  &  $0.8_{-0.31}^{+0.43}$  &  $1.04_{-0.12}^{+0.09}$  &  -                         \\ 
	UGC7323 &  $0.30_{-0.00}^{+0.02}$  &  $0.70_{-0.00}^{+0.01}$  &   $0.34_{-0.04}^{+0.06}$  &  $0.78_{-0.06}^{+0.02}$  &  $0.54_{-0.24}^{+0.03}$  &  $0.97_{-0.24}^{+0.01}$  &  $-11.60_{-1.30}^{+0.74}$  &  $0.6_{-0.12}^{+0.19}$  &  $1.01_{-0.09}^{+0.08}$  &  -                         \\ 
	UGC7399 &  $1.84_{-0.06}^{+0.11}$  &  $1.30_{-0.01}^{+0.00}$  &   $3.16_{-0.10}^{+0.12}$  &  $1.30_{-0.00}^{+0.00}$  &  $3.16_{-0.08}^{+0.14}$  &  $1.30_{-0.01}^{+0.00}$  &  -                         &  $5.0_{-0.08}^{+0.00}$  &  $1.30_{-0.01}^{+0.00}$  &  -                         \\ 
	UGC7524 &  $0.30_{-0.00}^{+0.01}$  &  $0.70_{-0.00}^{+0.01}$  &   $0.53_{-0.05}^{+0.03}$  &  $0.70_{-0.00}^{+0.00}$  &  $1.52_{-0.15}^{+0.21}$  &  $0.88_{-0.18}^{+0.10}$  &  $-11.20_{-0.35}^{+0.41}$  &  $1.9_{-0.34}^{+0.28}$  &  $0.91_{-0.20}^{+0.03}$  &  $-11.70_{-0.72}^{+0.47}$  \\ 
	UGC7559 &  -                       &  $0.70_{-0.00}^{+0.01}$  &   -                       &  $0.70_{-0.00}^{+0.01}$  &  -                       &  $0.98_{-0.12}^{+0.10}$  &  $-11.40_{-0.21}^{+0.21}$  &  -                      &  $0.96_{-0.22}^{+0.00}$  &  $-12.50_{-4.27}^{+0.61}$  \\ 
	UGC7577 &  -                       &  $0.70_{-0.00}^{+0.01}$  &   -                       &  $0.70_{-0.00}^{+0.02}$  &  -                       &  $1.00_{-0.11}^{+0.09}$  &  $-11.00_{-0.27}^{+0.30}$  &  -                      &  $1.00_{-0.11}^{+0.11}$  &  $-12.00_{-0.29}^{+0.57}$  \\ 
	UGC7603 &  $0.30_{-0.00}^{+0.03}$  &  $0.73_{-0.03}^{+0.01}$  &   $0.30_{-0.00}^{+0.04}$  &  $0.91_{-0.05}^{+0.02}$  &  $0.30_{-0.00}^{+0.03}$  &  $0.91_{-0.05}^{+0.03}$  &  -                         &  $0.4_{-0.08}^{+0.09}$  &  $1.17_{-0.08}^{+0.05}$  &  -                         \\ 
	UGC8490 &  $0.39_{-0.05}^{+0.05}$  &  $1.11_{-0.03}^{+0.03}$  &   $0.48_{-0.03}^{+0.06}$  &  $1.30_{-0.01}^{+0.00}$  &  $0.48_{-0.03}^{+0.06}$  &  $1.30_{-0.01}^{+0.00}$  &  -                         &  $1.4_{-0.00}^{+0.70}$  &  $1.30_{-0.14}^{+0.00}$  &  -                         \\ 
	UGC9211 &  $0.97_{-0.53}^{+0.32}$  &  $0.71_{-0.01}^{+0.05}$  &   $1.17_{-0.55}^{+0.71}$  &  $0.85_{-0.08}^{+0.05}$  &  $1.42_{-0.70}^{+0.66}$  &  $0.98_{-0.19}^{+0.00}$  &  $-12.70_{-4.43}^{+0.80}$  &  $3.0_{-1.43}^{+0.56}$  &  $1.00_{-0.14}^{+0.03}$  &  $-12.70_{-5.78}^{+0.00}$  \\ 
	UGC11707&  $0.33_{-0.03}^{+0.04}$  &  $0.70_{-0.00}^{+0.01}$  &   $0.83_{-0.09}^{+0.07}$  &  $0.70_{-0.00}^{+0.01}$  &  $1.55_{-0.20}^{+0.17}$  &  $0.70_{-0.00}^{+0.01}$  &  $-12.50_{-0.11}^{+0.25}$  &  $3.9_{-0.79}^{+0.28}$  &  $0.71_{-0.01}^{+0.10}$  &  $-12.10_{-0.13}^{+0.60}$  \\ 
	UGC11861&  $0.97_{-0.04}^{+0.03}$  &  $0.70_{-0.00}^{+0.01}$  &   $1.41_{-0.08}^{+0.02}$  &  $0.70_{-0.00}^{+0.01}$  &  $1.76_{-0.23}^{+0.06}$  &  $0.70_{-0.00}^{+0.06}$  &  $-12.00_{-0.17}^{+0.53}$  &  $2.5_{-0.25}^{+0.27}$  &  $0.97_{-0.13}^{+0.06}$  &  $-11.30_{-0.39}^{+0.26}$  \\ 
	UGC12060&  $0.98_{-0.13}^{+0.12}$  &  $0.70_{-0.00}^{+0.01}$  &   $1.69_{-0.24}^{+0.13}$  &  $0.70_{-0.00}^{+0.02}$  &  $2.66_{-0.40}^{+0.54}$  &  $0.98_{-0.12}^{+0.10}$  &  $-11.20_{-0.26}^{+0.29}$  &  $4.9_{-1.28}^{+0.07}$  &  $1.00_{-0.11}^{+0.09}$  &  $-10.80_{-0.47}^{+0.06}$  \\ 
	UGC12632&  $1.24_{-0.16}^{+0.12}$  &  $0.70_{-0.00}^{+0.01}$  &   $2.49_{-0.43}^{+0.07}$  &  $0.71_{-0.01}^{+0.03}$  &  $2.47_{-0.44}^{+0.13}$  &  $0.71_{-0.01}^{+0.03}$  &  -                         &  $5.0_{-0.78}^{+0.01}$  &  $0.95_{-0.17}^{+0.07}$  &  $-11.80_{-0.99}^{+0.38}$  \\ 
	F568-V1 &  $2.43_{-0.39}^{+0.30}$  &  $0.75_{-0.05}^{+0.03}$  &   $2.70_{-0.40}^{+0.51}$  &  $0.88_{-0.08}^{+0.06}$  &  $2.66_{-0.44}^{+0.44}$  &  $0.94_{-0.12}^{+0.02}$  &  -                         &  $5.0_{-0.78}^{+0.00}$  &  $1.04_{-0.13}^{+0.03}$  &  $-12.10_{-5.18}^{+0.45}$  \\ 
	F574-1  &  $1.81_{-0.11}^{+0.08}$  &  $0.70_{-0.00}^{+0.01}$  &   $2.50_{-0.15}^{+0.11}$  &  $0.70_{-0.00}^{+0.01}$  &  $5.00_{-0.23}^{+0.00}$  &  $0.94_{-0.08}^{+0.04}$  &  $-10.30_{-0.17}^{+0.00}$  &  $5.0_{-0.80}^{+0.00}$  &  $0.98_{-0.17}^{+0.05}$  &  $-11.10_{-0.81}^{+0.21}$  \\ 
	F583-1  &  $1.15_{-0.24}^{+0.13}$  &  $0.70_{-0.00}^{+0.03}$  &   $1.64_{-0.36}^{+0.27}$  &  $0.75_{-0.05}^{+0.03}$  &  $1.84_{-0.37}^{+0.31}$  &  $0.96_{-0.23}^{+0.01}$  &  $-12.20_{-1.11}^{+0.54}$  &  $2.3_{-0.46}^{+0.50}$  &  $0.97_{-0.11}^{+0.00}$  &  -                         \\ 
	F583-4  &  $0.41_{-0.11}^{+0.15}$  &  $0.88_{-0.10}^{+0.04}$  &   $0.69_{-0.21}^{+0.30}$  &  $0.96_{-0.11}^{+0.07}$  &  $1.21_{-0.81}^{+0.03}$  &  $1.00_{-0.14}^{+0.06}$  &  $-12.50_{-5.21}^{+2.06}$  &  $3.3_{-2.03}^{+0.00}$  &  $1.00_{-0.13}^{+0.08}$  &  $-11.70_{-6.46}^{+0.76}$  \\
	\hline
	\end{tabular}
\end{table*}

\section{Solar System constraints}
As shown in the previous section, emergent gravity does not produce high-quality fits to galaxy rotation curves, even in the most favorable case. Nevertheless, this could be due to the fact that the formula derived in spherical symmetry differs significantly from the one that should be applied in an axisymmetric case. We should thus also turn our attention to a regime in which this formula is fully applicable. The Solar System is a good test case since, as a first approximation, it is reasonable to consider the Sun as a spherically symmetric body. Moreover, because planets are outside of the bulk of matter, the derivative term in the transition function naturally cancels out, and Eq.~(\ref{eq:mond}) with the transition function given by Eq.~(\ref{eq:nuEG}) leads to 
\begin{equation}
	g=\frac{GM_\odot}{r^2}+\frac{\sqrt{a_MGM_\odot}}{r}\, ,
\end{equation}
with $r$ the distance to the Sun. Considering a two-body problem, the additional acceleration due to emergent gravity will produce (amongst others) a time evolution of the argument of perihelion $\omega$ given by the solution of 
\begin{equation}
	\frac{d\omega}{dt}=\sqrt{a_MGM_\odot}\frac{\sqrt{1-e^2}}{nae}\frac{\cos f}{r}\, ,
\end{equation}
with $a$ the semi-major axis, $e$ the eccentricity, $n=2\pi/P$ the mean motion ($P$ the period) and $f$ the true anomaly. At first order, one finds the expression of the secular advance of perihelion produced by emergent gravity (see also Eq. (64) from \cite{blome:2010kq} for a similar calculation)
\begin{equation}\label{eq:dw}
	\Big<\frac{d\omega}{dt}\Big>=\sqrt{\frac{a_M}{a}}\,\frac{1-e^2-\sqrt{1-e^2}}{e^2}\, .
\end{equation}

The motion of planets around the Sun is inferred by planetary ephemerides analyses from an impressive number of different observations: radioscience observations of spacecraft that orbited around Mercury, Venus, Mars and Saturn, flyby tracking of spacecraft close to Mercury, Jupiter, Uranus and Neptune and optical observations of all planets \cite{folkner:2009fk,*konopliv:2011dq,*folkner:2010kx,*folkner:2014uq,hees:2014jk,fienga:2008fk,*fienga:2009kx,*verma:2014jk,*fienga:2015rm,fienga:2011qf,pitjeva:2005kx,*pitjeva:2014fj,pitjeva:2013fk,*pitjev:2013qv}. Estimations of anomalous supplementary advances of perihelia have been derived with the INPOP (Int\'egrateur Num\'erique Plan\'etaire de l'Observatoire de Paris) ephemerides \citep{fienga:2011qf} and are given in Tab.~\ref{tab:sol}. These values correspond to  the interval in which the differences of postfit residuals in the ephemerides analysis are below 5 \%. Similar results have also been obtained by the Ephemerides of Planets and the Moon (EPM) \citep{pitjeva:2013fk,*pitjev:2013qv}. 

The secular advance of the argument of perihelion for the different planets produced by emergent gravity is seven orders of magnitude larger than what is currently allowed by observations, an unacceptable discrepancy. This is shown in Tab.~\ref{tab:sol} where the value of the advance of perihelion due to emergent gravity using an acceleration scale of $a_M$=6.9 $\times  10^{-11}$ m/s$^2$ (determined from the fit to galactic rotation curves in the most favorable case, see previous section) is presented. Larger values of $a_M$ would lead to an even larger discrepancy. Let us note that an hypothetical Galactic external field effect would add a quadrupole term to the problem, which is a severe constraint for MOND transition functions \cite{hees:2016mi,milgrom:2009vn,blanchet:2011ys,hees:2014jk}, but would not reduce the reported discrepancy for emergent gravity.  Such a difference, already mentioned in \citep{milgrom:2016aa}, is unacceptable and such an effect would  have undoubtedly been  observed. 

\begin{table}
\caption{Col. 2: estimation of the secular advance of the argument of perihelion $\omega$ for the different planets for the emergent gravity (see Eq. (\ref{eq:dw}) with $a_M$=6.9 $\times  10^{-11}$ m/s$^2$.  Col. 3: estimation of a secular advance of periastron from the INPOP planetary ephemerides (from \cite{fienga:2011qf}). Emergent gravity predicts effects on the planetary motion that are seven orders of magnitude too large compared to observations.}
\label{tab:sol} 
\centering
\begin{tabular}{l|rr}
& $\Big<\frac{d\omega}{dt}\Big>$ from Eq. (\ref{eq:dw})  & $\Big<\frac{d\omega}{dt}\Big>$ from \cite{fienga:2011qf} \\
&  mas/cy&   mas/cy\\\hline
Mercury    &  -1.01 $\times 10^7$   &  $0.4 \pm 0.6$      \\ 
Venus      &  -0.82 $\times 10^7$   &  $0.2\pm 1.5 $      \\ 
Earth      &  -0.70 $\times 10^7$   &  $-0.2\pm 0.9$      \\ 
Mars       &  -0.57 $\times 10^7$   &  $-0.04\pm 0.15$    \\ 
Jupiter    &  -0.31 $\times 10^7$   &  $-41\pm 42$        \\
Saturn     &  -0.23 $\times 10^7$   &  $0.15\pm 0.65$     \\ 
\hline
\end{tabular}
\end{table}

\section{Conclusions}
The emergent gravity developed by \citet{verlinde:2016qf} leads to predictions similar to the MOND phenomenology, but not quite identical. In this paper, we study its impacts on galactic rotation curves and within the Solar System. At galactic scales, the emergent gravity formula produces fits to rotation curves that are less satisfactory than using MOND, a conclusion also reached in a similar analysis \citep{lelli:2017aa}. First of all, the Hubble constant that is demanded by the data is extremely low compared to that inferred from the cosmic microwave background or other measurements (42.6 km/s/Mpc, see Sec.~\ref{sec:noprior}). Moreover, the preferred distances are systematically low, and the best-fit stellar mass-to-light ratios tend to be rather low too. And even by letting all these parameters free, and with the help of a putative external field effect, the fits produce reduced chi-square significantly higher than the MOND ones. This is due to the fact that emergent gravity produces wiggles in the rotation curves that are unobserved in the data. We caution that these conclusions rely on the assumption that the transition between the Newtonian and emergent gravity regimes derived in spherical symmetry is roughly applicable to axisymmetric systems. Nevertheless, since the problem is rooted in the dependence on the radial derivative of the Newtonian acceleration, we suspect this generic problem would not go away in a more rigorous axisymmetric case. The general lesson from this is that the MOND formula, whatever its origin, tends to produce a quite reasonable description of the observed data in rotationally-supported galaxies, and that a force law deviating from it is not guaranteed to be as successful, as illustrated here.

We then tested the predicted weak-field transition from emergent gravity in a regime where it should a priori be fully applicable, i.e. in the Solar System. There, emergent gravity produces a deviation from Newtonian gravity seven orders of magnitude larger than what is allowed by current measurements.  This rules out the present emergent gravity weak-field formula with high confidence, unless a screening-like mechanism can be found to reduce strongly this deviation in the Solar System. One example of such a mechanism is actually proposed in \citep{liu:2016uq} where a scenario of a field-like dark mass is suggested, with maximally anisotropic pressure. As noticed in \citep{liu:2016uq,iorio:2016aa}, this leads to deviations much smaller in the Solar System. Nevertheless, this scenario might lead to a Catch-22 problem since the MOND-like behavior is completely suppressed (this can be noticed in Eq.~(14) from \citep{liu:2016uq} where the logarithmic term responsible for the MOND-like behavior disappears in the case $w'=-1$ which corresponds to a field-like dark mass) and is therefore not viable either to explain galaxy rotation curves. 

The two problems addressed in this communication may hypothetically be solved by altering the current predictions from emergent gravity. For instance, a mechanism could be found to reduce the expected deviation from General relativity in the Solar System. Nevertheless, the nature of the two problems which we uncovered is very different: the discrepancy at galactic scales is rooted in the dependence on the radial derivative of the Newtonian acceleration, while this derivative has no impact in the Solar System predictions. This fundamental difference may be an obstacle for modifying the current formalism in order to fit both types of observations.  In the absence of an alteration of the present formalism, we conclude that the emergent gravity as developed in \citep{verlinde:2016qf} is not viable observationally.



\bibliography{emergent}

\appendix 

\end{document}